\documentclass[%
reprint,
superscriptaddress,
 amsmath,amssymb,
 aps,
prb,
]{revtex4-2}

\usepackage{graphicx}
\usepackage{dcolumn}
\usepackage{bm}
\usepackage{svg}
\usepackage{caption}
\usepackage{subcaption}
\usepackage{xcolor}
\usepackage{adjustbox}  
\usepackage{orcidlink}
\newcommand{\etal}{{\it et al.}}

\newcommand{\sgn}{{\hbox{\small sgn}}}

\newcommand{\eref}[1]{Eq.~(\ref{#1})}
\newcommand{\fref}[1]{Fig.~\ref{#1}}


\usepackage{hyperref}
\usepackage{bibunits}  
\defaultbibliography{refs,refs_mine,references,supp}

\begin{document}
\begin{bibunit}[apsrev4-2]

\preprint{APS/123-QED}

\title{Universal transport at 
 Lifshitz metal-insulator transitions in two dimensions 
}%

\author{Harry Tomlins\orcidlink{0009-0000-4172-971X}}
\affiliation{Department of Physics, King’s College London, Strand, London WC2R 2LS, United Kingdom}
\author{Jan M.\ Tomczak\,\orcidlink{0000-0003-1581-8799}}
\affiliation{Department of Physics, King’s College London, Strand, London WC2R 2LS, United Kingdom}
\affiliation{Institute of Solid State Physics, TU Wien, 1040 Vienna, Austria}

\date{\today}

\begin{abstract}
We study the charge transport across a band-tuned metal-insulator transition in two dimensions.
For high temperatures $T$ and chemical potentials $\mu$ far from the transition point, conduction is ballistic and the resistance $R(T)$ verifies a simple one-parameter scaling relation.
Here, we explore the limits of this semi-classical behaviour and study the quantum regime beyond, where 
scaling breaks down. We derive an analytical formula for the simplest Feynman diagram of the linear-response conductivity $\sigma=1/R$ 
of a parabolic band endowed with a finite lifetime. 
Our formula shows excellent agreement for experiments for a field-tuned MoTe$_2$/WSe$_2$ moiré bilayer, and can capture the quantum effects responsible for breaking the one-parameter scaling.
We go on to discuss a fascinating prediction of our model: 
The resistance at the quantum-critical band-tuned Lifshitz point ($\mu=T=0$) has the {\it universal value}, $R_L=(2 \pi h)/e^2$, per degree of freedom and this value is found to be compatible with experiment. Furthermore, we investigate whether two dimensional metal-insulator transitions driven by strong electron correlations or disorder can also be classified by their quantum-critical resistance and find that $R_L$ may be useful in identifying predominantly interaction driven transitions.

\end{abstract}

\maketitle


Metal-insulator transitions (MIT) in two dimensional materials have been studied intensely (and controversially) for decades \cite{Kravchenko_2004,app9061169,SARMA2005579,RevModPhys.73.251}. 
It is consensus that the phenomenon involves a formidable interplay of disorder and electronic correlations.
Hence, when a change in carrier density triggers a transition, pinpointing its microscopic driver---strong electron-electron interactions leading to Mott-Wigner localisation \cite{mott,Camjayi2008,PhysRevB.82.155102,PhysRevB.106.155145},
strong disorder causing Anderson localisation \cite{PhysRev.109.1492,RevModPhys.57.287},
or a more conventional band-like transition \cite{Lifshitz1960} at intermediate correlations and disorder---remains far from trivial.
An observable
directly probing the metal-insulator transition is the resistance  and much can be learnt from its temperature profile $R(T)$ \cite{RevModPhys.82.1743}.
Approximating charge carriers as ballistic, a semi-classical treatment of conduction \cite{Universal_Scal} predicts that
$R(T)$ verifies a simple one-parameter scaling law \cite{RevModPhys.73.251,Kravchenko_2004}:
$R(T)=R_c(T)\times\mathcal{F}_{\sgn(\mu)}(T_0(\mu)/T)$. Here, $R_c(T)$ is the critical resistance profile at the transition, that we shall identify with the origin of the chemical potential, $\mu=0$. Then, when scaling temperature $T$ with the $\mu$-dependent parameter $T_0(\mu)$, the ratio $R(T)/R_c(T)$ will collapse onto one of two universal curves, $\mathcal{F}_{\sgn(\mu)}$, distinguishing whether the system is a metal ($\mu>0$: $\mathcal{F}_+$) or an insulator ($\mu<0$: $\mathcal{F}_-$).
The gist of this mathematical statement is that, while conduction in the metal and in the insulator will be different ($\mathcal{F}_{+}\neq\mathcal{F}_{-}$), both regimes are controlled by a {\it single}, common energy-scale, $T_0(\mu)$.
In traditional 2D systems displaying an MIT---foremost metal-oxide-semiconductor field-effect transistors \cite{Kravchenko_MOSFET,PhysRevLett.87.266402} (MOSFET) and 2D hetero-structure quantum-well (QW) devices \cite{PhysRevB.99.081106,PhysRevLett.80.1288,COLERIDGE2000268,PhysRevB.72.081313,PhysRevB.57.R15068}---deviations from this scaling behaviour are experimentally nigh absent.

\begin{figure}[h!]
    \centering
    \includegraphics[width=0.9\linewidth]{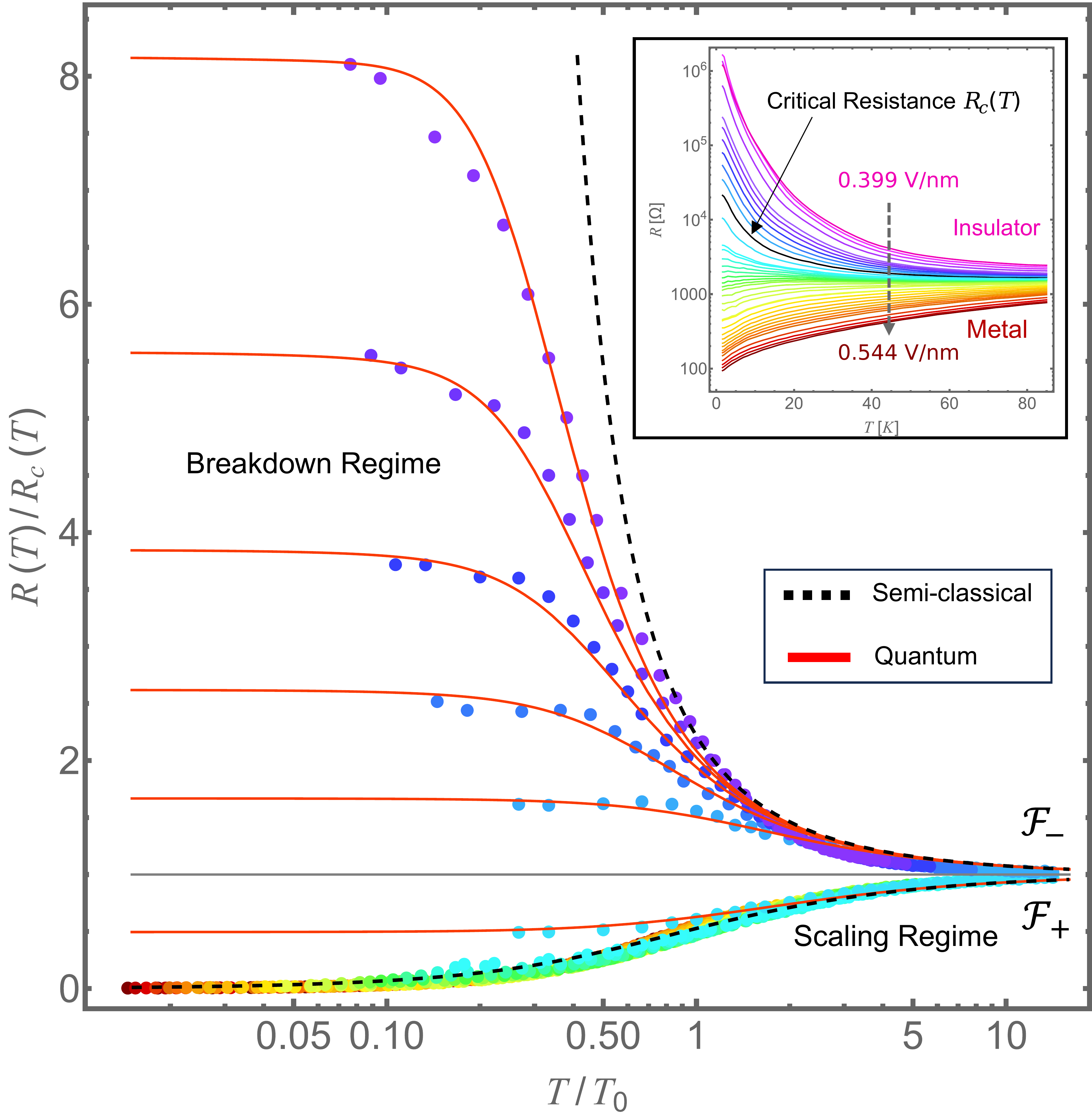}
    \caption{{\bf Metal-insulator transition in a MoTe$_2$/WSe$_2$ moiré bilayer.} Tuning an applied electric displacement field from $D=$0.544~V/nm (red dots) to $D=$0.427~V/nm (purple dots), the MoTe$_2$/WSe$_2$ bilayer near full  filling ($f=2$) undergoes a metal-insulator transition. The inset shows the data from the experiment by Li {\it et al.}~\cite{MoTe2_data} provided in Ref.~\cite{Disorder_Dom_Crit}, which is then scaled to give the data points in the main figure.
    In the high-temperature scaling regime, resistance curves $R(T)$ collapse onto a metallic or insulating branch 
    (black dashed lines $\mathcal{F}_+$ and $\mathcal{F}_-$)
    when dividing resistances $R(T)$ by the critical resistance $R_c(T)$ (shown in black in the inset, corresponding to $D=$0.437~V/nm) and scaling temperature with a field-dependent $T_0$ (see Suppl. Materials). Nearing the quantum critical point, $R_c(T=0)$, the experimental data strongly deviates from the semi-classical prediction ($\mathcal{F}_+$ and $\mathcal{F}_-$) that verifies one-parameter scaling, while agreeing excellently with our quantum theory (red solid lines) for all temperatures.} 
    \label{experimental_scaling}
\end{figure}

However, the recent emergence of intrinsically two-dimensional systems, such as graphene, transition-metal dichalcogenides (TMDs) and the twisted heterostructures assembled from them, has provided a new platform 
to study
2D physics with unprecedented precision \cite{Moon2021,PhysRevB.84.205325,CHOI2017116}. Crucially, metal-insulator transitions
can now be perused continuously via an applied electrical field.
In \fref{experimental_scaling} we display a key experiment \cite{MoTe2_data,Disorder_Dom_Crit}:
Adding two holes to a MoTe$_2$/WSe$_2$ moiré bilayer via gating depletes its first valence band (filling factor $f=2$) and then by applying an external  electrical displacement field the band-width of the first and second moiré valence band is tuned, manipulating the gap or overlap between the bands, leading to a band-insulator to metal transition of the Lifshitz type.
The resistances across that transition (coloured dots in \fref{experimental_scaling}) verify the above {\it one}-parameter scaling (dashed lines) in a regime limited to high temperatures and sufficiently deep into the metal phase.  
For $T/T_0\lesssim 2$ the measured resistances no longer collapse onto the semi-classical prediction and an ill-understood non-universal behaviour emerges---suggesting that {\it multiple} energy scales are at play. 

In this work, we develop an analytical theory “as simple as possible, but no simpler” that describes conduction across and arbitrarily close to the MIT.
The theory's results (red solid lines in \fref{experimental_scaling}) encompass the semi-classical high-$T$ limit (itself "too simple"), while also providing an excellent description of the experimental near-critical conduction in the quantum regime beyond. 
From the vantage point of our theory, we take a fresh look at 2D metal-insulator transitions. 
Our analysis reveals: (1) At low temperatures, the electronic scattering rate $\Gamma$ turns into a relevant energy scale, invalidating the one-parameter scaling theory.
(2) Strikingly, we find the transport of any band-tuned system at the Liftshitz point, $R(\mu=0,T\rightarrow 0)=R_L=(2 \pi h)/e^2$ to be {\it universal}, in congruence
with the ($f=2$) MoTe$_2$/WSe$_2$ bilayer experiment.
Finally, (3), we discuss more broadly how, in different types of MITs in 2D, the 
quantum critical resistance may be related to the underlying transition mechanism. We motivate that $R_L$ provides a sufficient but not necessary condition for identifying transitions driven predominately by interactions.

\section{Modelling the phase transition} 

We model the metal-insulator transition in a 2D system by a single parabolic band, $\epsilon_{\mathbf k}=\frac{\hbar^2 k^2}{2m}$, with the reduced Planck constant $\hbar$, the band-mass $m$, and Brillouin-zone momentum ${\mathbf k}$.
This 2D electron gas (2DEG) dispersion is realisable in MOSFETs and quantum-well devices and approximates well the transport-dominating
band-edges in van-der-Waals semiconductors, such as TMDs.
Adjusting the position of the chemical potential, $\mu$, the system can then be 
tuned from a metal ($\mu>0$) to a band insulator ($\mu<0$). Moving through the critical value, $\mu=0$, where the chemical potential is touching the band edge, the system undergoes a Lifshitz transition as the Fermi surface disappears \cite{Lifshitz1960,Topological_liftshitz_transition,QPT_from_topology_in_momentum_space}. 
To allow for quantitative comparisons with specific materials, we will multiply the conductance with an integer factor $N_d$ to account for the electron's spin, the presence of both valence and conduction bands, band degeneracies and the possible occurrence of multiple valleys in the Brillouin zone. 
We model electronic scattering by endowing the dispersion with a finite lifetime $\tau$, which can depend on temperature and, in principle, on the chemical potential. 
The finite scattering rate $\Gamma=\hbar/(2\tau)$ effectively mimics the effects of disorder and electronic correlations.
We, however, make no attempt at describing its precise microscopic origin, which may vary between systems. 
Deliberately assuming the scattering rate as static, further precludes 
capturing metal-insulator transitions of the Mott-Wigner type. 

For this electronic structure,
we derive an exact expression for the leading-order longitudinal linear-response electrical conductance $\sigma$, per spin, band and valley degree of freedom.
The corresponding Feynman diagram (that neglects vertex-corrections) yields
\cite{Coleman_2015,LRT_prototyp}
\begin{equation}
    \sigma(T)=\frac{he^2}{2V}\sum_{\mathbf{k}}\int_{-\infty}^\infty d\omega \left(-\frac{\partial f}{\partial \omega}\right) \left(A_{\mathbf{k}}^{\phantom{x}}(\omega)v_{\mathbf{k}}^x\right)^2
    \label{kubo}
\end{equation}
with the electron's charge $e$, Planck's constant $h$, the unit-cell volume $V$  the Fermi function $f(\omega)$, and the Fermi velocity $v_{\mathbf{k}}^x=\hbar^{-1}\partial/\partial_{k_x}\epsilon_{\mathbf{k}}$ in $x$-direction.
For the static scattering rate considered, the spectral function has a Lorentzian line-shape: $A_{\mathbf{k}}(\omega)=-1/\pi \operatorname{Im}(\omega+\mu-\epsilon_{\mathbf{k}}-i\Gamma)^{-1}$. 
Then, the $\omega$-integral in \eref{kubo} can be evaluated analytically using contour-integration techniques \cite{LRT_Tstar,LRT_prototyp,LRT,jmt_fesb2}. In the current case, also the momentum summation can be performed explicitly, leading to our central result
%
    \begin{align}
     \sigma( \mu,T)= &\frac{e^2}{2\pi h}\left\{\frac{\pi k_B T}{\Gamma} \ln2\pi +\frac{\pi}{2}\frac{\mu}{\Gamma}\right.\nonumber\\      
 &+\operatorname{Re}\Psi_0\left(\frac{1}{2}+\frac{1}{2\pi (k_B T)}(\Gamma-i \mu)\right)\label{cond}\\
 &\left. -\frac{2\pi k_B T}{\Gamma}\operatorname{Re}\ln\boldsymbol\Gamma\left(\frac{1}{2}+\frac{1}{2\pi k_B T}(\Gamma-i \mu)\right)\right\},\nonumber
    \end{align}
%
where $\ln \boldsymbol\Gamma(z)$ is the natural logarithm of the Gamma function, $\Psi_0(z)=d/dz \ln\boldsymbol\Gamma(z)$ is the digamma function \cite{Abramowitz}
and $k_B$ Boltzmann's constant. 
This exact expression for the conductance without vertex-corrections makes no assumption on the magnitude of scattering: It is valid for transport in the ballistic and the diffusive limit, as well as in between. The effective mass of the band does not occur explicitly in our expression, which allows for a general application to any 2D parabolic band system. 

For context, we revisit the ballistic, semi-classical limit, 
previously derived by Fratini \etal\ \cite{Universal_Scal}. Considering $T/\Gamma$ and $|\mu|/\Gamma\gg 1$,  \eref{cond} reduces to (see Supplementary Material)
\begin{equation}
    \sigma^{sc}(\mu,T)=\frac{e^2}{2 h}\left(\frac{k_BT}{\Gamma}\log\left[1+e^{\frac{\mu}{k_BT}}\right]\right).
    \label{condsc}
\end{equation}
Manifestly, the approximate resistance, $R^{sc}=1/\sigma^{sc}$, now obeys the scaling relation $R^{sc}(\mu,T)=R^{sc}(0,T)\times\mathcal{F}_{\sgn(\mu)}({\mu}/{T})$ with $\mathcal{F}_{\hbox{\small sgn}(x)}(x)=1/\log_2\left[1+e^{{\sgn(x)}/{x}}\right]$.

\section{Metal-insulator transition in a Moiré bilayer}


%
%
To validate our theory, we model the experimental data of the field-tuned MIT in the MoTe$_2$/WSe$_2$ bilayer near full-filling ($f=2$) \cite{MoTe2_data} (see Suppl. Material), shown in \fref{experimental_scaling}  (coloured dots) with \eref{cond} (red lines). Following Fratini \etal\ \cite{Universal_Scal} we make the ansatz, $\Gamma(T)=\Gamma_0+b T$,
for the scattering rate.
Using $\Gamma_0=0.3$meV and $b=0.005$meV/K we find outstanding agreement with the scaled experimental data.
We clearly see in \fref{experimental_scaling} that,
for large (scaled) temperatures, $T/T_0\gtrsim 2$, the above branches $\mathcal{F}_+$ and $\mathcal{F}_-$ (black dashed lines)
describe the data of, both, the metallic and the insulating phase accurately, as found previously \cite{Disorder_Dom_Crit,Universal_Scal}.
At lower temperatures and close to the Lifshitz transition, however, the behaviour is qualitatively different:
The excellent agreement with \eref{cond} for experimentally available temperatures suggests that in the metal ($\mu>0$), instead of vanishing, the scaled resistance, $R(T)/R_c(T)$, extrapolates to a finite residual value for $T\rightarrow 0$. 
In the insulator, instead of diverging as predicted semi-classically, the scaled resistance also saturates.

\section{Transport in the Quantum Regime}

\begin{figure*}
        \begin{subfigure}{0.31\textwidth}
        \centering
        \includegraphics[scale=0.073]{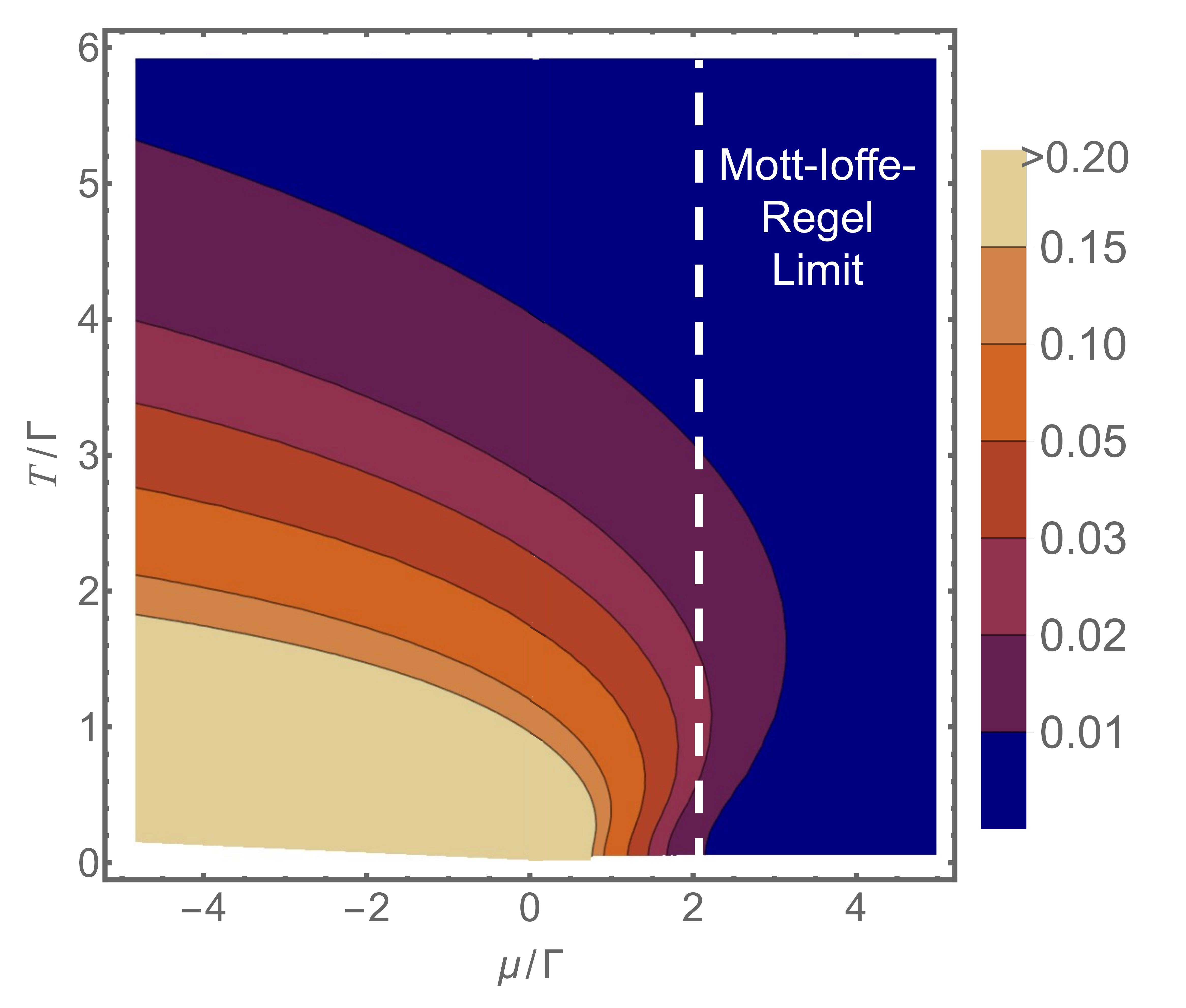} 
            \caption{Deviations of the quantum resistance formula from its semi-classical limit, $\left|1-{R^{sc}(\mu,T)}/{R(\mu,T)}\right|$}
             \label{semiclassical-validity}
    \end{subfigure}
    \begin{subfigure}{0.31\textwidth}
        \centering
        \includegraphics[scale=0.07]{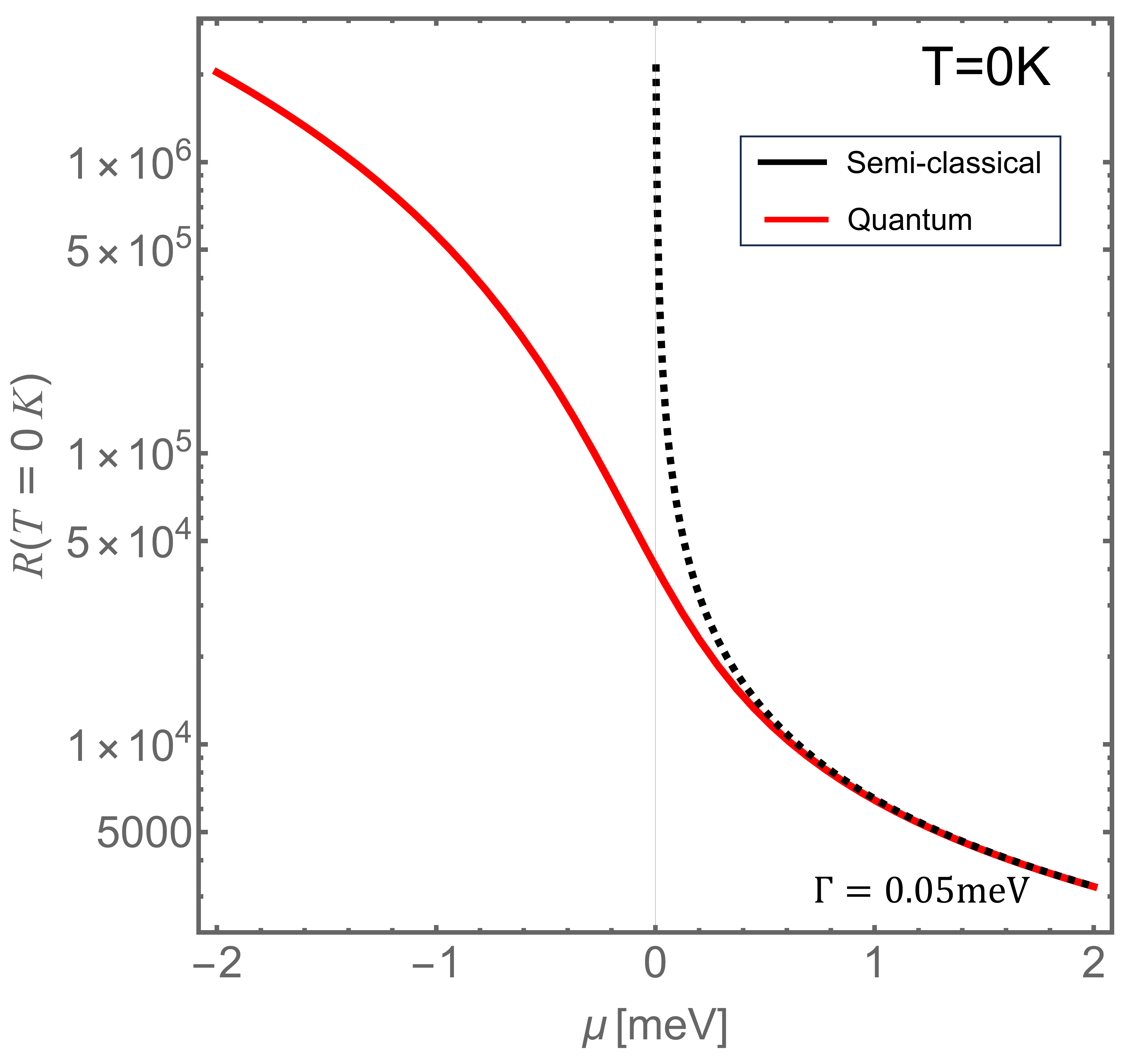}  
            \caption{Quantum vs.\ semi-classical transport at $T=0K$\newline \newline}
             \label{Fig3b}
    \end{subfigure}
    \begin{subfigure}{0.31\textwidth}
        \centering
        \includegraphics[scale=0.07]{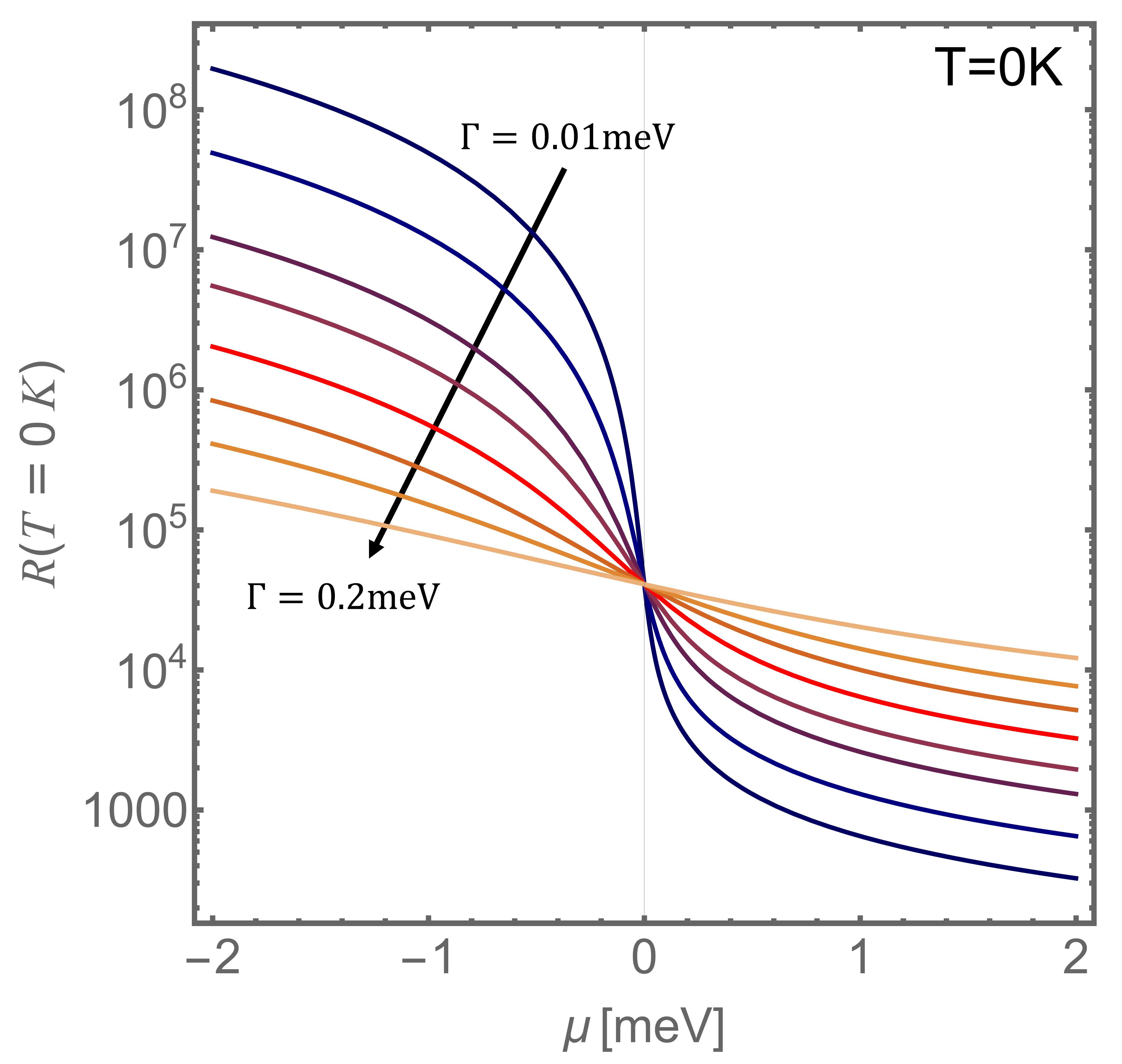}  
            \caption{Effect of the scattering rate $\Gamma$ on the resistance $R(\mu,0)$ at $T=0K$\newline \newline}
             \label{universal_point}
    \end{subfigure}
    \caption{\textbf{Quantifying transport in the quantum regime}. (a) The relative deviation of the quantum resistance $R(\mu,T)$ of \eref{cond}, from the semi-classical limit, $\left|1-{R^{sc}(\mu,T)}/{R(\mu,T)}\right|$, for a constant scattering rate $\Gamma$ is shown. 
   At low temperatures close to the transition we see a breakdown of semi-classical behaviour in the metallic phase, corresponding, broadly, to the Mott-Ioffe-Regel limit.
    The semi-classical approximation is accurate deep in the metal, but, also, for arbitrary chemical potential, when temperature is high enough. (b) Semi-classically the resistance in the insulating phase at $T=0$K is predicted to diverge, however in the quantum theory with the inclusion of finite lifetime effects, a finite value is seen for all values of $\mu$.  (c) The residual resistance is shown for an increasing value of the scattering rate $\Gamma$. At $\mu=0$ all the curves intersect, indicating a universal value for the resistance at the Lifshitz quantum critical point.}
\end{figure*}

\subsection{Limits of a semi-classical description} 

Deviations from semi-classical scaling in the quantum regime imply that,
besides the chemical potential $\mu$, the scattering rate $\Gamma$ has become a relevant energy scale, with major consequences for the charge transport. We start by quantifying the quantum regime by calculating the relative deviation
of the full quantum resistance formula to its semi-classical limit, $|1-R^{sc}/R|$, in \fref{semiclassical-validity}.
Under the assumption of a constant scattering rate, both,  \eref{cond} and \eref{condsc} only depend on the two dimensionless ratios $\mu/\Gamma$ and $k_BT/\Gamma$. But only in the approximate conductance, \eref{condsc}, do these dependencies factorise, allowing for scaling: $\sigma^{sc}(\mu,T)=\sigma^{sc}(0,T/\Gamma)/\mathcal{F}(\mu/T)$.
Instead, in the much richer exact conductance, \eref{cond}, the interplay of $\mu/\Gamma$ and $k_BT/\Gamma$ can be non-trivial and highly non-linear. 

We observe that deep into the metallic phase (${\mu}/{\Gamma}\gtrsim$2) the semi-classical expression provides an excellent approximation (blue intensity = negligible deviations) all the way down to $T=0$. However, as the system approaches the band-tuned Lifshitz transition, deviations start to occur at low temperatures. This finding is consistent with the phenomenological Mott-Ioffe-Regel criterion \cite{IR_limit,MIR_limit},
which states that for transport in lattice systems to be ballistic, the distance $l$ between scattering events cannot be smaller than the inter-atomic separation. In reciprocal space, this requirement is commonly expressed as $k_Fl\gtrsim 1$, with the Fermi wave vector $k_F$.
Estimating the mean-free path as $l=v_{k_F}\tau$ via the group velocity $v_{k_F}$ and the lifetime $\tau=\hbar/(2\Gamma)$, one finds for the electron gas $k_Fl=\mu/(2\Gamma)\gtrsim 1$, which is indicated by the dashed vertical line in \fref{semiclassical-validity}.

On the insulating side, deviations from semi-classical behaviour are much more pronounced:
There, the lifetime-broadening included in the exact response leads to a much higher conductance than predicted in the semi-classical picture which only includes thermal (not lifetime) broadening \cite{LRT_Tstar}.


\subsection{ Transport at zero temperature}

The effect of lifetime-broadening is most striking at $T=0$ where thermal broadening is absent.
Namely our new theory suggests that the residual ($T\rightarrow 0$) conductance changes {\it continuously} through the field-tuned metal-insulator transition. An asymptotic expansion of \eref{cond} yields
\begin{equation}
    \sigma(\mu,0)=\frac{e^2}{2\pi h}\left[1+\frac{\mu}{\Gamma}\left(\frac{\pi}{2}+\arctan\left(\frac{\mu}{\Gamma}\right)\right)\right].
    \label{residual}
\end{equation}
In line with the experiment we find it most instructive to plot the resistance (1/$\sigma$) and as shown in \fref{Fig3b}, the residual resistance is indeed a uniformly continuous (and differentiable) function of the chemical potential.
As a consequence, unlike what is predicted semi-classically (dashed line), the resistance at $T=0$ remains finite in the insulating phase. The microscopic origin of this resistance saturation is the presence of lifetime-broadened, incoherent spectral weight in the gap \cite{LRT_prototyp}, despite the absence of thermal broadening, as discussed previously for correlated narrow-gap semiconductors \cite{NGCS,LRT_Tstar}. This effect becomes larger with increasing $\Gamma$ as shown in \fref{universal_point}.


\section{Universal Transport at the Quantum Critical Point}

We now combine some of the above understanding to reveal a fascinating feature of the 2D band-tuned Lifshitz transition.
As seen in \fref{universal_point}, for different scattering rates the theoretical residual resistance $R(\mu,T=0)$ has an isosbestic point: meaning at the critical point $\mu=k_BT=0$ the resistance is independent of the scattering rate. 
In fact, the resistance is independent of any parameter (carrier density or effective mass) of the system and has (per spin, band and valley degree of freedom)  the universal value

\begin{equation}
    R_L=(2\pi h)/e^2.
    \label{unicond}
\end{equation}

We will refer to $R_L$ as the `Liftshitz resistance'. This result implies that for any 2D system which undergoes a band-tuned (``BT'') metal-insulator transition, the resistance at their quantum critical point, $R_c^{BT}=R(\mu=0,T=0)$, should take the same unique value $R_L$, 
even across materials with different scattering mechanisms.
To verify this prediction, we confront it with the experimental data from the ($f=2$)  MoTe$_2$/WSe$_2$ bilayer system \cite{MoTe2_data}.
Since experiments are performed at finite temperatures (the experimental base temperature is $T=1.6$K) and for a discrete set of displacement fields only, we assess the compatibility of our prediction by extracting a confidence interval for the experimental critical resistance $R
_c^{exp}(T\rightarrow 0)$ (see Suppl.\ Materials for details).
We find an estimated range of
$ 0.3\, R_L\leq R_c^{exp} \leq 1.2\, R_L $
(per degree of freedom of the system), 
meaning that our prediction of universal quantum critical transport is compatible with the experiment of Li \etal\ \cite{MoTe2_data}.
This encouraging result calls for future measurements. 
To reduce the experimental error bar for the MoTe$_2$/WSe$_2$ bilayer system, i.e.,  further approaching its Lifshitz quantum-critical point, reaching lower temperatures and sampling a finer set of displacement fields is required. Also, more systems exhibiting an in principle continuously band-tunable MIT for which our prediction could be tested need to be identified.

  \begin{figure*}
      \centering

  \begin{subfigure}[c]{0.48\textwidth}
        \centering
        \includegraphics[scale=0.3]{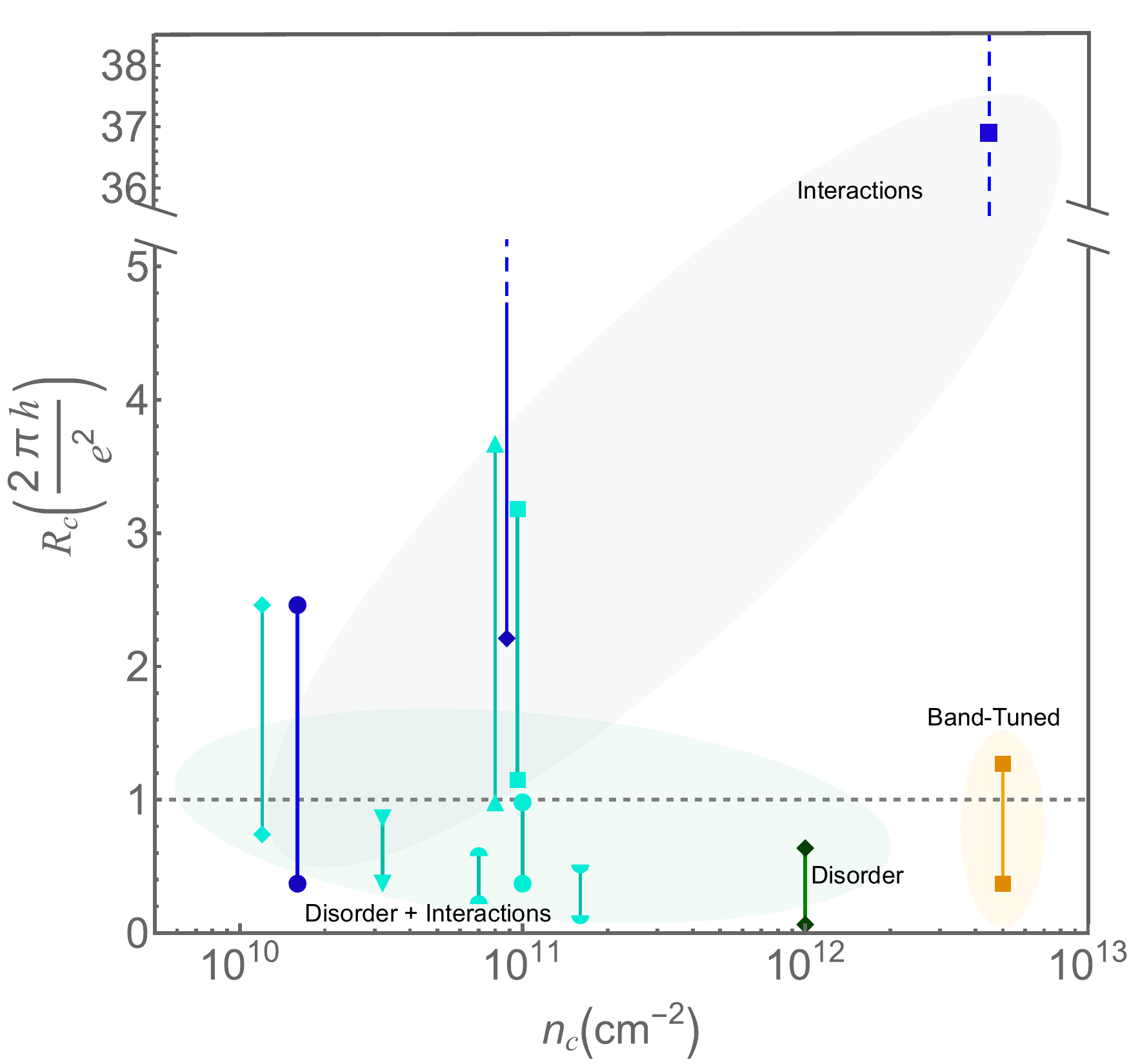}
    \end{subfigure}%
    \begin{subfigure}[c]{0.48\textwidth}
        \centering
        \includegraphics[scale=0.23]{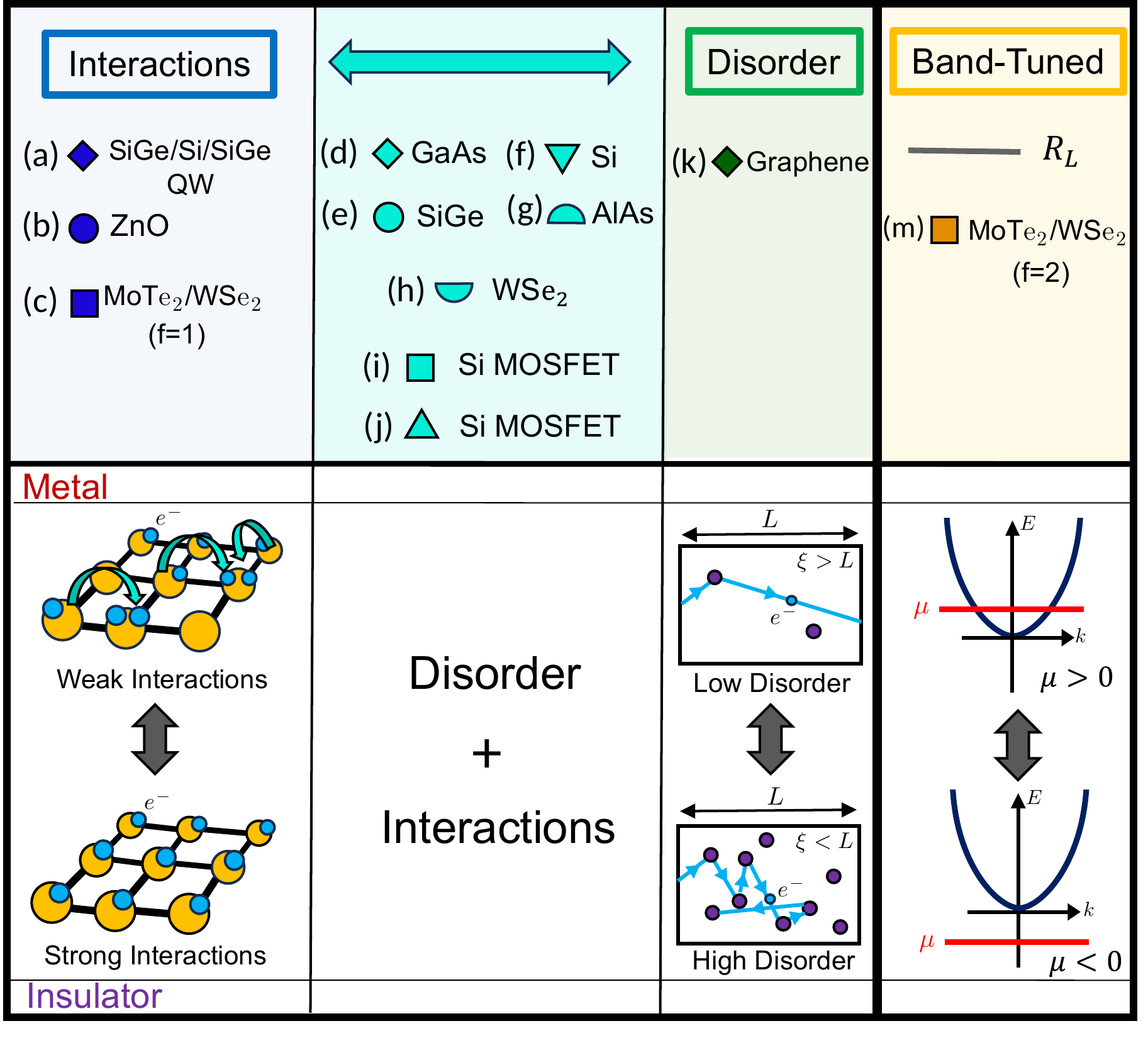} 
    \end{subfigure}
   \caption{\textbf{Comparison of experimental critical resistance ranges with $\boldsymbol{R_L}$.} Extracted critical resistance ranges from several experiments (per degree of freedom of each system), plotted against the carrier density $n_c$ at which the system was critical. The systems are labelled by the predominant mechanism which drives the MIT. In the intermediate regime where this assignment is unclear, systems are labelled ``disorder + interactions''. The Lifshitz resistance $R_L$, the predicted quantum critical resistance for a band-tuned transition and proposed as an upper bound for a predominantly disorder-driven transition sets the scale of the resistance and is shown as a horizontal grey dashed line. Dotted vertical lines indicate that the range extends beyond the limits of the graph. A sensible critical resistance range for data point (c), the half-filled ($f=1$) MoTe$_2$/WSe$_2$ bilayer could not be estimated using our standard method for the data available (hence the dashed line). However the transition takes place near the point indicated.
  Experimental references of systems shown: (a) \cite{PhysRevB.99.081106}, (b) \cite{Falson2022}, (c,m) \cite{MoTe2_data}, (d) \cite{PhysRevLett.80.1288}, (e) \cite{COLERIDGE2000268}, (f) \cite{PhysRevB.72.081313}, (g) \cite{PhysRevB.57.R15068}, (h) \cite{Pack2024}, (i) \cite{Kravchenko_MOSFET}, (j) \cite{PhysRevLett.87.266402}, (k) \cite{Osofsky2016}.
  }
    \label{critical_res} 
    \end{figure*}
\subsection{Identifying the transition mechanism of a MIT using $R_c$}
A natural question to ask is whether the value of $R_c^{BT}=R_L$ 
can be used to distinguish band-tuned transitions from MITs driven by other mechanisms.
More precisely, is there a relation to the critical resistances, $R_c^{D}$ and $R_c^{I}$, of transitions dominantly driven by disorder (``D'') or interactions (``I'')?

The behaviour of charge carriers in many real 2D electronic systems is governed by a complex interplay of both disorder and interactions. Determining which is more dominant in driving the MIT can not always be identified experimentally, especially in experiments where the MIT is induced by changing the carrier density on a sample-by-sample basis \cite{PhysRevLett.82.1744,app9010080,Brussarski2018,app8101909,Pack2024}. Therefore, a simple identifying feature, such as $R_c$, could be a very useful tool in helping to understand the complex physics behind a given 2D MIT. 

In highly disordered systems, the low-energy electronic states may undergo Anderson localisation and the system becomes insulating when the chemical potential drops below the mobility edge $\mu_c$ \cite{Anderson_textbook,Mott_Localisation,PhysRev.109.1492}. 
We expect our theory for the residual conduction, \eref{residual}, to apply for $\mu\geq \mu_c\geq0$, that is, in the metallic phase above the mobility edge.
Since the resistance drops monotonously with decreasing chemical potential,
$R_c^{D}=R(\mu=\mu_c,T=0)<R(\mu=0,T=0)=R_L$,
we can hypothesise that the Lifshitz resistance $R_L$ provides an \textit{upper bound} for the critical resistance, $R^{D}_c<R_L$, of a predominately disorder-driven MIT. This statement again assumes that vertex corrections can be neglected. Indeed, to first approximation, ladder-type of vertex corrections can be accounted for by a renormalised scattering rate (see, e.g., Ref.~\cite{Coleman_2015}), which then has no influence on $R_L$, which is independent of $\Gamma$. In turn, weak localisation effects have been shown to be suppressed near the MIT in several relevant 2D systems \cite{Kravchenko_2004,PhysRevLett.91.116402}.

However for a predominately interaction-driven MIT, a similar constraint on $R_c^{I}$ cannot be defined. In fact, 
at a Mott-Wigner transition the electronic effective mass diverges and this is accompanied by a strongly dynamical scattering rate, $\Gamma(\omega)$, that engenders a vanishing quasi-particle lifetime on the former Fermi surface \cite{Camjayi2008,PhysRevB.106.155145}. The approximation of a static $\Gamma$, made in the derivation of \eref{cond}, is thus no longer justified. Hence, our theory can provide no indication on the value of $R^{I}_c$.

Therefore we propose a sufficient but not necessary 
criterion;
that if $R_c>R_L$ then this signifies the predominant influence of interactions in a 2D metal-insulator transition. However, for $R_c\leq R_L$ nothing conclusive about the origin of the MIT can be deduced. 
Indeed, we predict that all three types of transitions can occur at $R_c=R_L$, whilst both disorder and interaction driven MITs can also occur at values $R_c<R_L$.

\subsection{Investigating the $R_c$ criterion}

To explore this criterion
we extracted the `critical resistance ranges' (per degree of freedom) from a variety of 2D systems that were driven through a MIT.  Ideally each system could be categorised by the predominant driver of the MIT. However, in many systems due to the highly non-trivial interplay between interactions and disorder no conclusive cause for the MIT has previously been identified. Where clear prior evidence for one mechanism was found, the appropriate label is given, otherwise the system is labelled as `disorder + interaction' driven.

The `critical resistance ranges' plotted in Figure \ref{critical_res} show broadly the expected pattern, that whilst transitions driven predominately by interactions (blue symbols) span a huge range of $R_c$ values, transitions where disorder should also play an important role in the MIT lie around or below the value of $R_L$ (cyan symbols).

The systems with the largest $R_c$ values (points (a) SiGe/Si/SiGe QW \cite{PhysRevB.99.081106} and (c) ($f=1$) MoTe$_2$/WSe$_2$ \cite{MoTe2_data} in Fig.~\ref{critical_res} occurring at $R_c\sim 10R_L$ and $R_c\sim 37R_L$ respectively) display transitions that are very predominantly interaction driven, with the electronic effective mass in both systems diverging before the transition \cite{PhysRevB.99.081106,MoTe2_data}. Therefore, we hypothesise that for $R_c\gg R_L$ the system will  be very strongly interaction driven, however the appearance of clear experimental signatures may make our criterion 
less necessary for such identification.
It is in the intermediate regime $R_c\sim R_L$ where interactions and disorder are on a more equal footing that our  
criterion
may be most useful in identifying when interactions are more relevant for the transition. For example, our 
criterion
supports the proposal \cite{Kravchenko_2004} that despite contributions from disorder, it is electronic interactions that 
are the protagonists 
in driving the transition of the two Si MOSFET devices (points (i) \cite{Kravchenko_MOSFET} and (j) \cite{PhysRevLett.87.266402} in \fref{critical_res}), as their critical resistance ranges ($R_c>R_L$) lie above the upper bound for a predominantly disordered transition. For the remaining intermediate `disorder + interaction' systems our criterion
is unable to conclusively classify the transitions. Having established that $R_c$ can be a useful tool for identifying interaction-driven 2D MITs in some situations,  further exploration is warranted into how the value of $R_c$ for an arbitrary system will be affected as the relative strength of disorder and electronic correlations are adjusted, experimentally as well as in theoretical descriptions of, e.g., the Anderson-Hubbard model.

\section*{Acknowledgements}
This research was funded
in part by the Austrian Science Fund (FWF) [\href{https://doi.org/10.55776/I6142}{10.55776/I6142}].
For open access purposes, the authors have applied a CC BY public copyright license to any author-accepted manuscript version arising from this submission.

\putbib
\end{bibunit}

\bigskip
\newpage

\begin{bibunit}[apsrev4-2]
\appendix
\setcounter{figure}{0}
\onecolumngrid
\newpage

\section*{Supplementary Information}
\section{Modelling the metal-insulator transition in a MoTe$_2$/WSe$_2$ bilayer}
\begin{figure*}[!htb]
    \centering
    \begin{subfigure}{0.3\textwidth}
        \centering
        \includegraphics[scale=0.15]{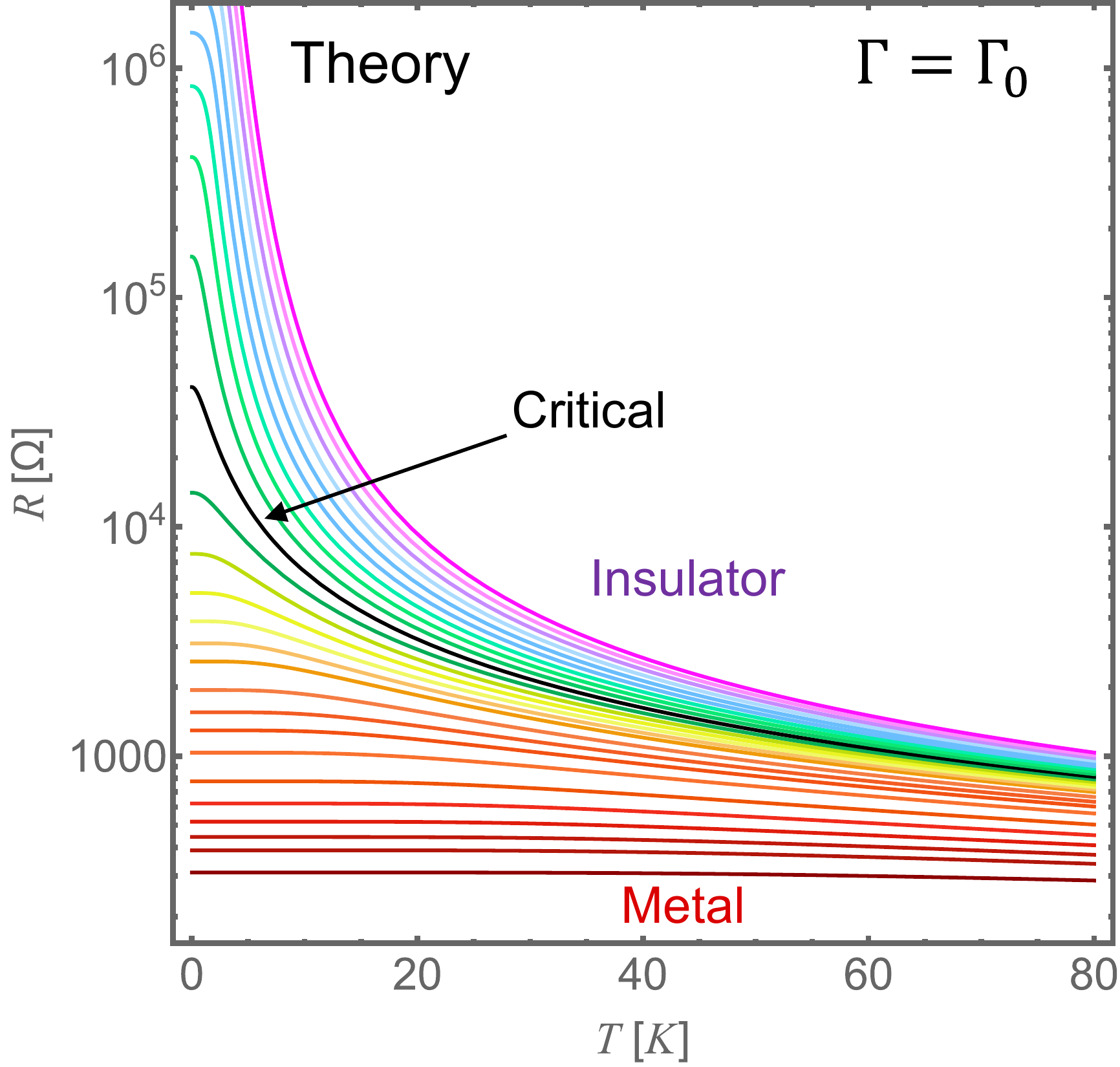}
        \caption{Resistance $R(T)$ for constant scattering $\Gamma=$ $\Gamma_0$=constant}
        \label{fig:prob1_6_2}
    \end{subfigure}%
    \begin{subfigure}{0.3\textwidth}
        \centering
        \includegraphics[scale=0.15]{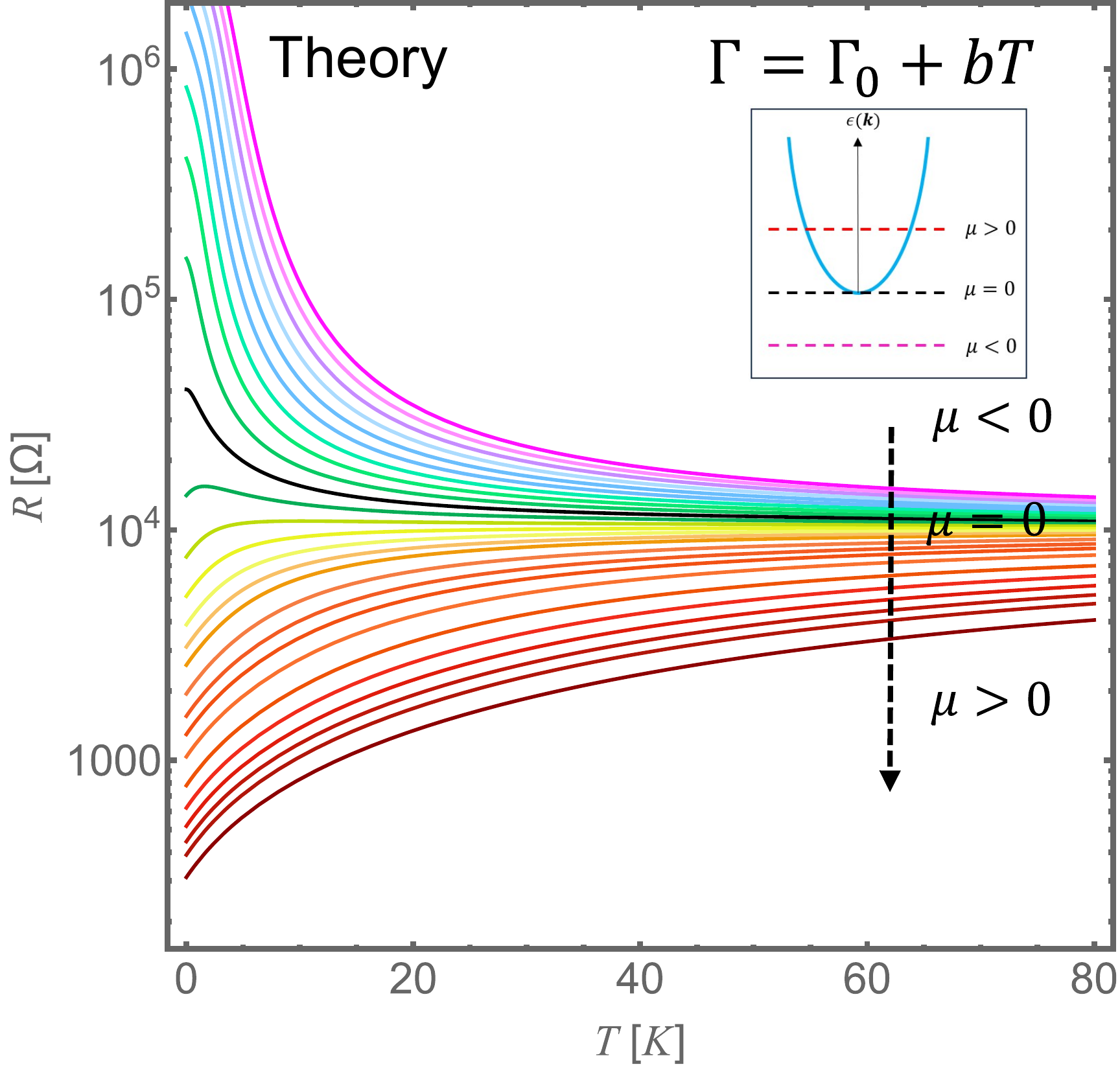}
        \caption{Resistance $R(T)$ for a scattering rate $\Gamma=$ $\Gamma_0$+ $bT$}
        \label{fig:prob1_6_1}
    \end{subfigure}
    \begin{subfigure}{0.3\textwidth}
        \centering
        \includegraphics[scale=0.15]{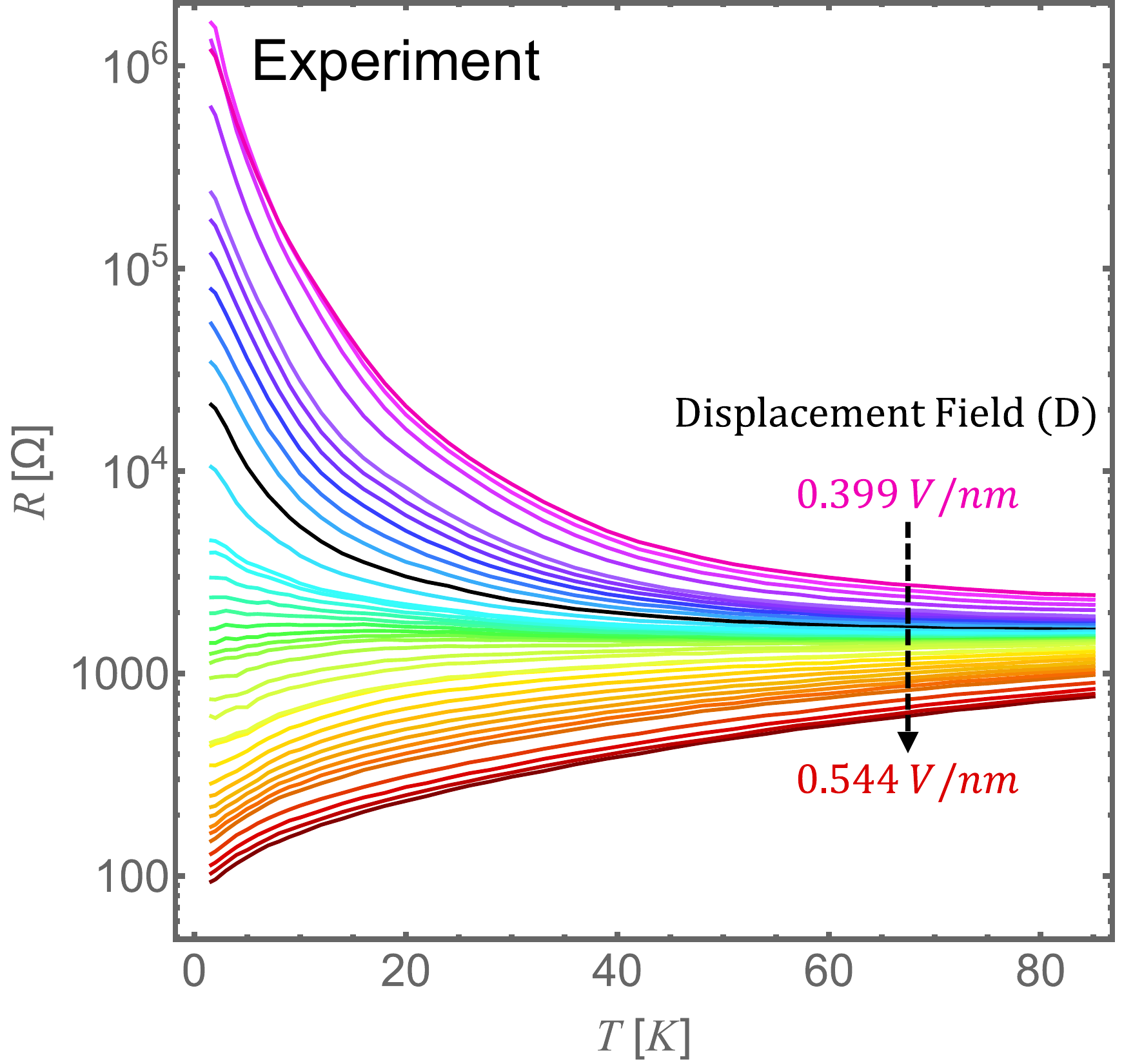}
        \caption{Experimental resistance curves of  a MoTe$_2$/WSe$_2$ bilayer }
        \label{fig:prob1_6_1}
    \end{subfigure}

    \caption{\textbf{Qualitative fit to MoTe$_2$/WSe$\boldsymbol{_2}$ experiment}. Resistance curves, $R(T)$, for varying chemical potential $\mu$, with red curves in the metallic and purple curves in the insulating regime. The black lines indicate the 'critical' resistances at $\mu=0$. (a) and (b) are theoretical curves. In  (a) a constant scattering rate $\Gamma=\Gamma_0=0.3$meV is used and (b) with an additional linear term, $\Gamma=\Gamma_0+bT$ with $b=0.005$meV/K is used. 
    (c) experimental resistances for the ($f=2$) MoTe$_2$/WSe$_2$ bilayer undergoing a MIT (data from Ref.~\cite{Disorder_Dom_Crit} of experiment carried out in Ref.~\cite{MoTe2_data}), each curve corresponding to a different applied electric displacement field $D$.  
}
\label{generated data}
\end{figure*}
\begin{figure}[h!]
    \centering
    \includegraphics[scale=0.25]{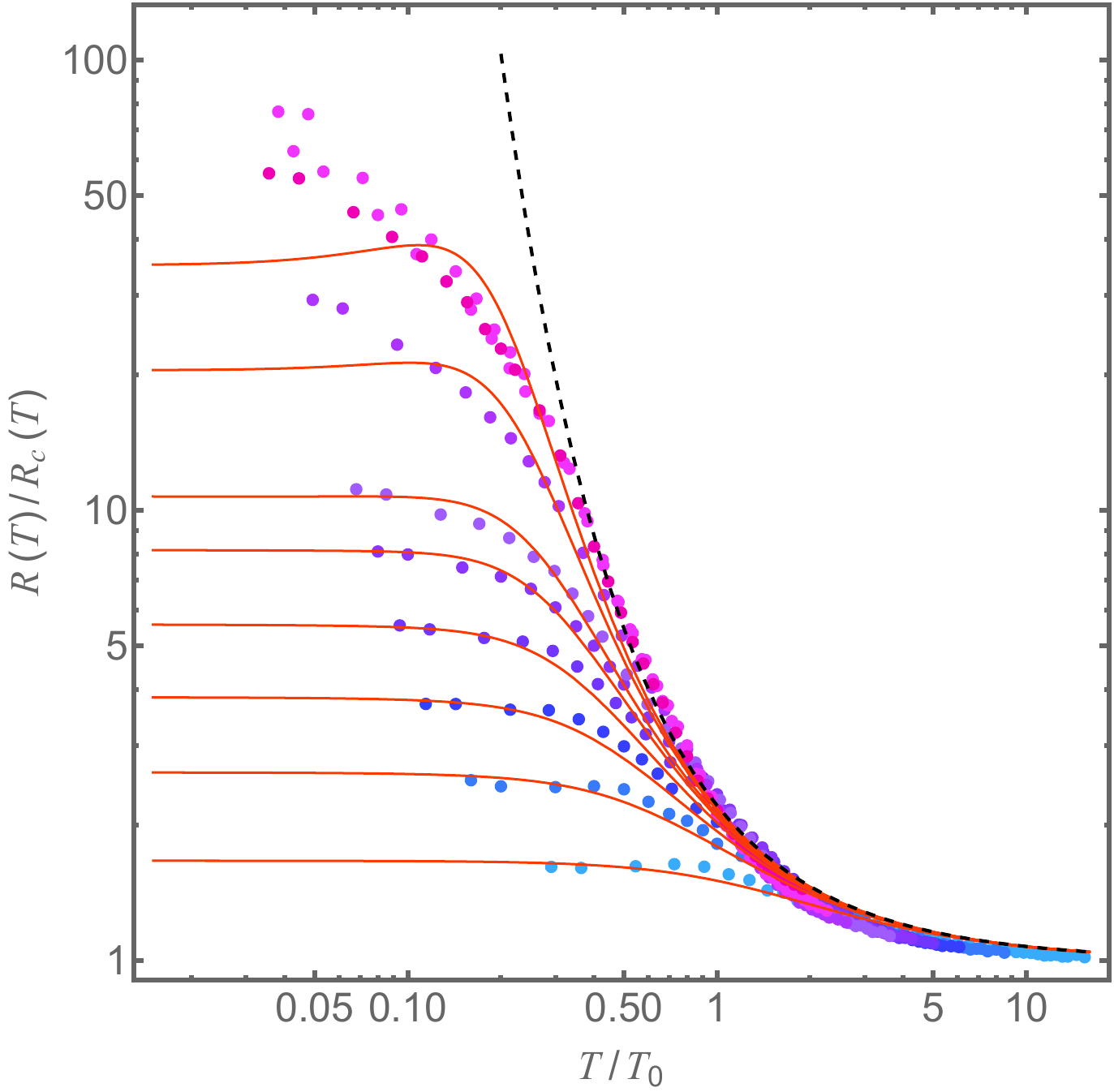}
    \caption{\textbf{Deviations from the quantum theory for larger gapped insulating states} \textbf{at low temperatures}. The scaled resistance curves for an applied electric displacement field $D=$0.399V/nm-0.435V/nm are shown, corresponding to all insulating curves measured in the experiment. As in the main text the theoretical curves use a scattering rate, $\Gamma= \Gamma_0+bT$ with $\Gamma_0=0.3$ meV and $b=0.005$ meV/K, that is independent of the chemical potential. It can be seen that for the four most insulating curves at low temperatures there is a deviation of the scaled data from our theory, being more resistive than predicted, however the theory still performs well close to the transition. We note that relaxing the restriction of a $\mu$ independent $\Gamma$, may help to improve the agreement between theory and experiment. }
    \label{bad fit}
\end{figure}

\newpage
\section{Low Temperature Behaviour}

To investigate the low temperature behaviour of the conductance an asymptotic expansion is required, as the variable $z=\frac{1}{2\pi k_B T}(\Gamma-i \mu)\rightarrow \infty$ when $T\rightarrow0$.  Here we provide the relevant expressions \cite{prototypical_transport}
\begin{equation}
    \ln \Gamma\left(\frac{1}{2}+z\right)=\frac{1}{2}\ln 2\pi+ z\ln z -z -\frac{1}{24z} +\mathcal{O}(z^{-3})
\end{equation}
\begin{equation}
    \Psi_0\left(\frac{1}{2}+z\right)=\ln z +\frac{1}{24z^2}+\mathcal{O}(z^{-4})
\end{equation}
To proceed we will use the following identity
\begin{equation}
    \ln(a+ib)=\frac{1}{2}\ln(a^2+b^2)+i\arctan\left(\frac{b}{a}\right)
\end{equation}
Hence, to leading order we find the following expressions 
\begin{equation}
\begin{aligned}
    \operatorname{Re}\ln\Gamma\left(\frac{1}{2}+z\right)&=\frac{1}{2}\ln 2\pi-\frac{\mu}{2\pi k_B T}\arctan\left(\frac{\mu}{\Gamma}\right)\\
    &-\frac{\Gamma}{2\pi k_B T} +\frac{\Gamma}{2\pi k_B T} \ln\frac{1}{2\pi k_B T}\\
    &+\frac{\Gamma}{2\pi k_B  T}\frac{1}{2}\ln(\Gamma^2+\mu^2)
\end{aligned}
\end{equation}
\begin{equation}
    \begin{aligned}
        \operatorname{Re}\Psi_0\left(\frac{1}{2}+z\right)=\ln \frac{1}{2\pi T}+\frac{1}{2}\ln(\Gamma^2+\mu^2)
    \end{aligned}
\end{equation}
Substituting these expressions back into the conductivity, we recover the residual conductivity value as quoted in the main text, and the first term for small but finite temperatures:
\begin{equation}
    \sigma(\mu,T)=\frac{e^2}{2\pi h}\left(1+\frac{\mu}{\Gamma}\left(\frac{\pi}{2}+\arctan\left(\frac{\mu}{\Gamma}\right)\right)
    +\frac{(2\pi)^2}{12}\frac{1}{(1+(\mu/\Gamma)^2)^2}\left(\frac{k_B T}{\Gamma}\right)^2+O\left(({k_B T}/{\Gamma})^4\right)\right)
    \label{lowT}
\end{equation}

Note, a dependency similar to the residual ($T=0$) term in Eq.~\ref{lowT} was previously derived for the effect of coupling a metallic lead to a 2D material with a hybridization strength $\Gamma$ \cite{Hall_TaS2}.

\section{Semi-classical Approximation}

Here, the semi-classical limit of the full quantum conductance is derived, under the assumption that $\Gamma=\frac{1}{2\tau}=$const. 
First, we define the variables $T'=\frac{k_B T}{\Gamma}$ and $\mu'=\frac{\mu}{\Gamma}$ which allows the conductance to be rewritten as 
 \begin{equation}
    \begin{aligned}
 \sigma( \mu,T)= &\frac{e^2}{2\pi h}\left(\pi T' \ln2\pi +\frac{\pi}{2}\mu'\right.\\      
 &+\operatorname{Re}\Psi_0[\frac{1}{2}+\frac{1}{2\pi T'}(1-i \mu')]\\
 &\left.-2\pi T'\operatorname{Re}\ln\boldsymbol{\Gamma}[\frac{1}{2}+\frac{1}{2\pi T'}(1-i \mu')]\right)
    \end{aligned}
    \label{appendix cond}
 \end{equation}
Then we take the semi-classical limit $T'\gg 1$, $|\mu'|\gg 1$.  To proceed we Taylor expand each expression noting that $z=\frac{1}{2\pi T'}(1-i \mu')$ will be a small quantity.

\begin{equation}
\begin{aligned}
   & \Psi_0\left(\frac{1}{2}+z\right)\approx\Psi_0\left(\frac{1}{2}\right)+\Psi_1\left(\frac{1}{2}\right) z + \frac{1}{2}\Psi_2\left(\frac{1}{2}\right)z^2 + O\left(z^3\right)\\
   &\ln\boldsymbol{\Gamma}\left(\frac{1}{2}+z\right)\approx\ln\boldsymbol{\Gamma}\left(\frac{1}{2}\right)+\Psi_0\left(\frac{1}{2}\right) z + \frac{1}{2}\Psi_1\left(\frac{1}{2}\right)z^2+ O\left(z^3\right)
\end{aligned}
\end{equation}
Using these expansions (and keeping only terms up to $O\left(\frac{1}{T'}\right)$) the conductivity can be approximated as
\begin{equation}
\begin{aligned}
    \sigma(\mu,T)\approx &\frac{e^2}{2\pi h}\left(\pi T' \ln2\pi +\frac{\pi}{2}\mu'
    +\Psi_0\left(\frac{1}{2}\right)+ \Psi_1\left(\frac{1}{2}\right)\frac{1}{2\pi T'}\right.\\
    &\left.-\ln\Gamma\left(\frac{1}{2}\right) 2\pi T'-\Psi_0\left(\frac{1}{2}\right) - \frac{1}{2}\Psi_1\left(\frac{1}{2}\right)\frac{1}{2\pi T'}\left(1-\mu'^2\right) \right)
 \end{aligned}
\end{equation}
Then taking the relevant limits as outlined above this expression reduces to 
\begin{equation}
\begin{aligned}
    \sigma(\mu,T)\approx &\frac{e^2}{2\pi h}\left(2\pi T'\left(\frac{1}{2}\ln2\pi-\ln\Gamma\left(\frac{1}{2}\right)\right)\right.\\
    &\left.+\frac{\pi}{2}\mu'+\frac{1}{2}\Psi_1\left(\frac{1}{2}\right)\frac{1}{2\pi T'}\mu'^2\right)
\end{aligned}
\end{equation}
Substituting the expression back in terms of the original variables we find 
\begin{equation}
\begin{aligned}
    \sigma(\mu,T)\approx &\frac{e^2}{h} \frac{k_B T}{\Gamma} \left(\left(\frac{1}{2}\ln2\pi-\ln\Gamma\left(\frac{1}{2}\right)\right)\right.\\
    &\left.+\frac{1}{4}\frac{\mu}{k_B T} +\frac{1}{8\pi^2}\Psi_1\left(\frac{1}{2}\right)\left(\frac{\mu}{k_B T}\right)^2\right)
\end{aligned}
\end{equation}
Finally, assuming $\mu/T$ to be small
\begin{equation}
\begin{aligned}
    \sigma(\mu,T)\approx &\frac{e^2}{2h } \frac{k_B T}{\Gamma}  \left( \log2+\frac{1}{2}\frac{\mu}{k_B T} +\frac{1}{8}\left(\frac{\mu}{k_B T}\right)^2\right)\\
    &\approx \frac{e^2}{2h } \frac{k_B T}{\Gamma} \log[1+e^{\frac{\mu}{k_B T}}]
\end{aligned}
\end{equation}
Hence, we recover the semi-classical expression.

\newpage
\section{Scaling of Experimental Data}

In order to scale the experimental resistance curves, the 'critical' resistance curve $R_c(T)=R(\mu=0,T)$ and the corresponding chemical potential $\mu$ of all the other resistance curves is required.

The closest estimate of the critical resistance curve is found to correspond to $D=$0.437V/nm, which was identified by the authors in Ref.~\cite{Disorder_Dom_Crit} by extrapolating the vanishing activation energy in the insulating regime along with the disappearance of the (very) low-temperature leading linear-T regime.

Identifying the chemical potential $\mu$ provides a subtler challenge. In the experiment performed for the MoTe$_2$/WSe$_2$ bilayer \cite{MoTe2_data} the chemical potential was changed via an external applied electric field. Therefore, whilst we know that $\mu(D)$ will be monotonic on either side of the critical point, the exact functional relationship is unknown. Hence, when performing temperature scaling of each resistance curve a parameter $T_0(D)$ was used, which acts in the same way as the chemical potential.

\begin{figure}[h!]
    \centering
    \includegraphics[scale=0.25]{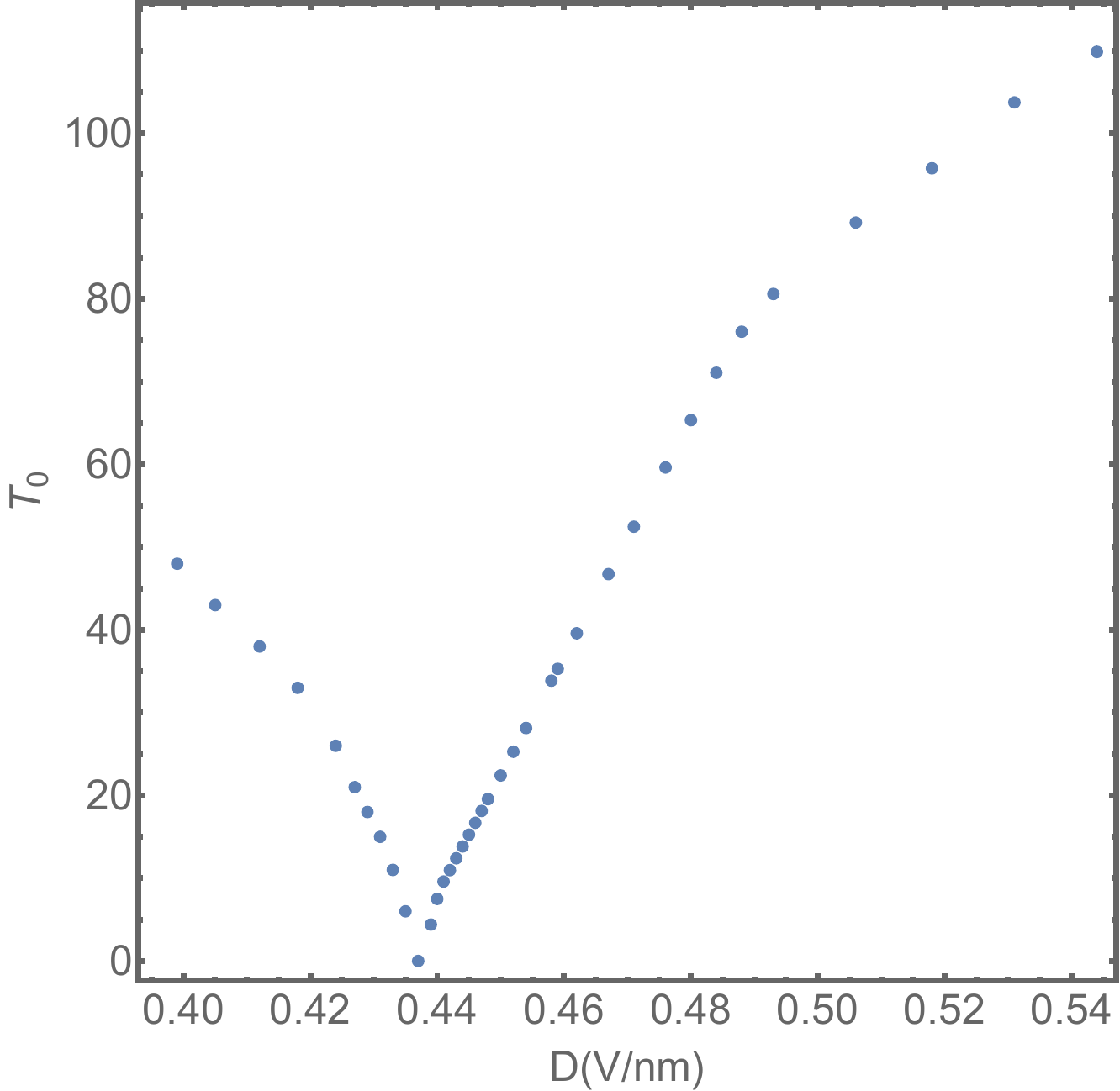}
    \caption{\textbf{Scaling parameter} $\boldsymbol{T_0}$. The values of the scaling parameter $T_0(D)$ as a function of applied electric 
 displacement field, used to scale the experimental data in the main text. The values ensure a collapse the resistance curves at high temperatures.}
    \label{T0_values}
\end{figure}

Figure \ref{T0_values} shows the values of $T_0(D)$ used to scale the experimental data in the main text. Each value of $T_0(D)$ was found by scaling the temperature axis of each experimental resistance curve such that, at high temperatures, the data collapsed onto the semi-classical scaling functions $\mathcal{F}_+$ and $\mathcal{F}_-$.

\newpage

\section{Estimating the $R_c$ range for the $(f=2)$ MoTe$_2$/WSe$_2$ System}

In order to conclusively verify our prediction of universal transport at the quantum critical point of a band-tuned MIT it is necessary to have an \textbf{accurate} value for $R_c(T=0K)$  from the experiment. However, from the data available this is not straight forward. Ideally, one could take the lowest temperature resistance measurement for the 'critical' ($D=$0.437 V/nm) resistance curve (as identified by \cite{Disorder_Dom_Crit,Universal_Scal}) and use that value as good approximation for the $T=0K$ resistance. However, the lowest temperature data point available is at $T=1.6K$ which is not low enough to act as a sensible approximation for the $T=0K$ value.
 Secondly, while the resistance curve measured at $D=0.437$ V/nm is identified as the 'critical' curve, due to the electric field resolution of the experiment, this may not correspond to the true $\mu=0$ critical point and ‘small’ deviations away from $\mu=0$ can have large impacts on the value at $T=0$K. 

\begin{figure}[h!]
    \centering
    \includegraphics[scale=0.2]{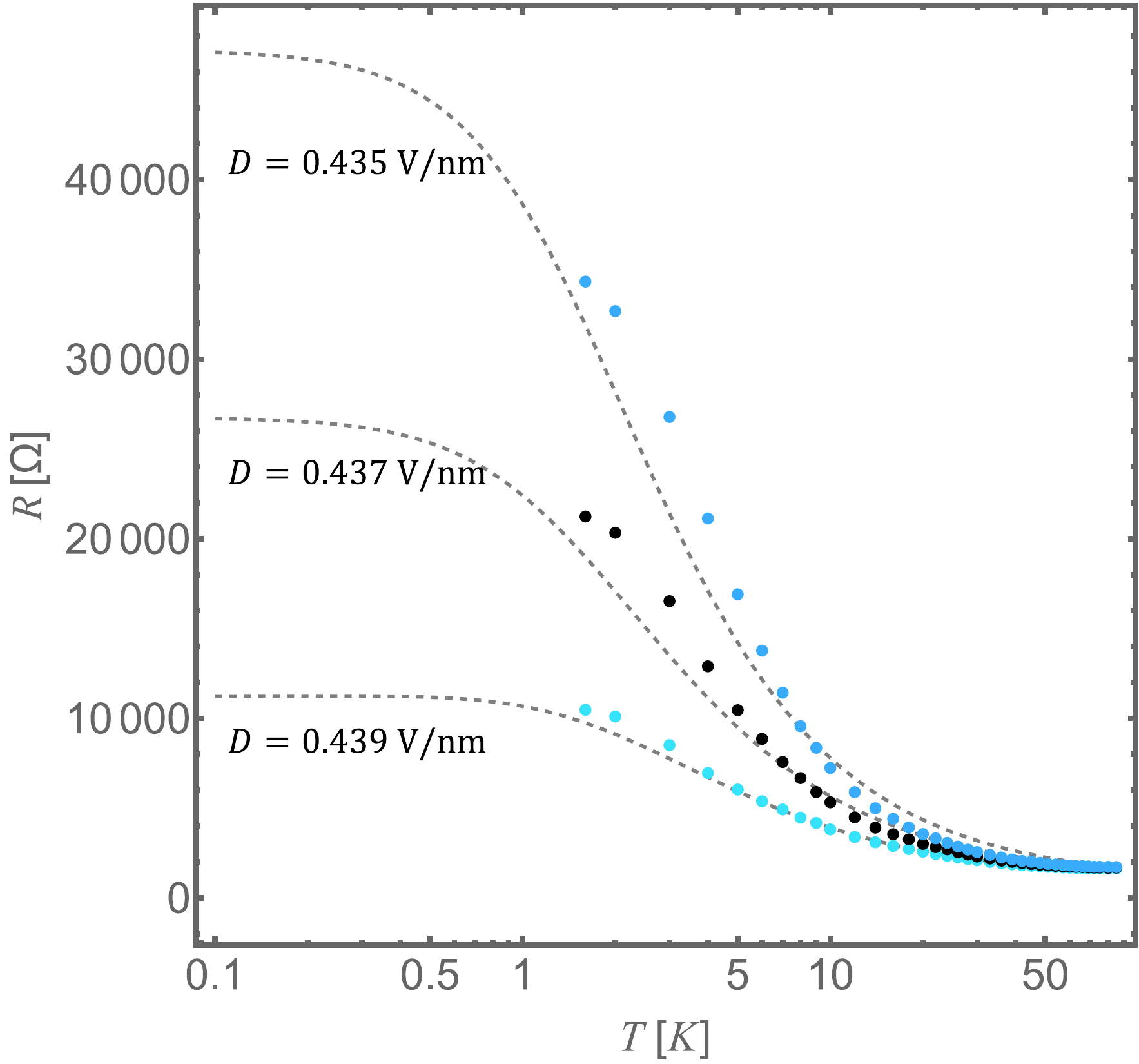}
    \caption{\textbf{Direct fitting of experimental resistance data.} In order to extrapolate the experimental data to T=0K we perform a best fit of our theory to the experimental data. We tune the chemical potential $\mu$, the residual scattering rate $\Gamma_0$ and the temperature coefficient $b$ in $\Gamma(T)=\Gamma_0+bT$ to provide a fit to the data.}
    \label{direct_fit}
\end{figure}

\begin{figure}[h!]
    \centering
    \includegraphics[scale=0.2]{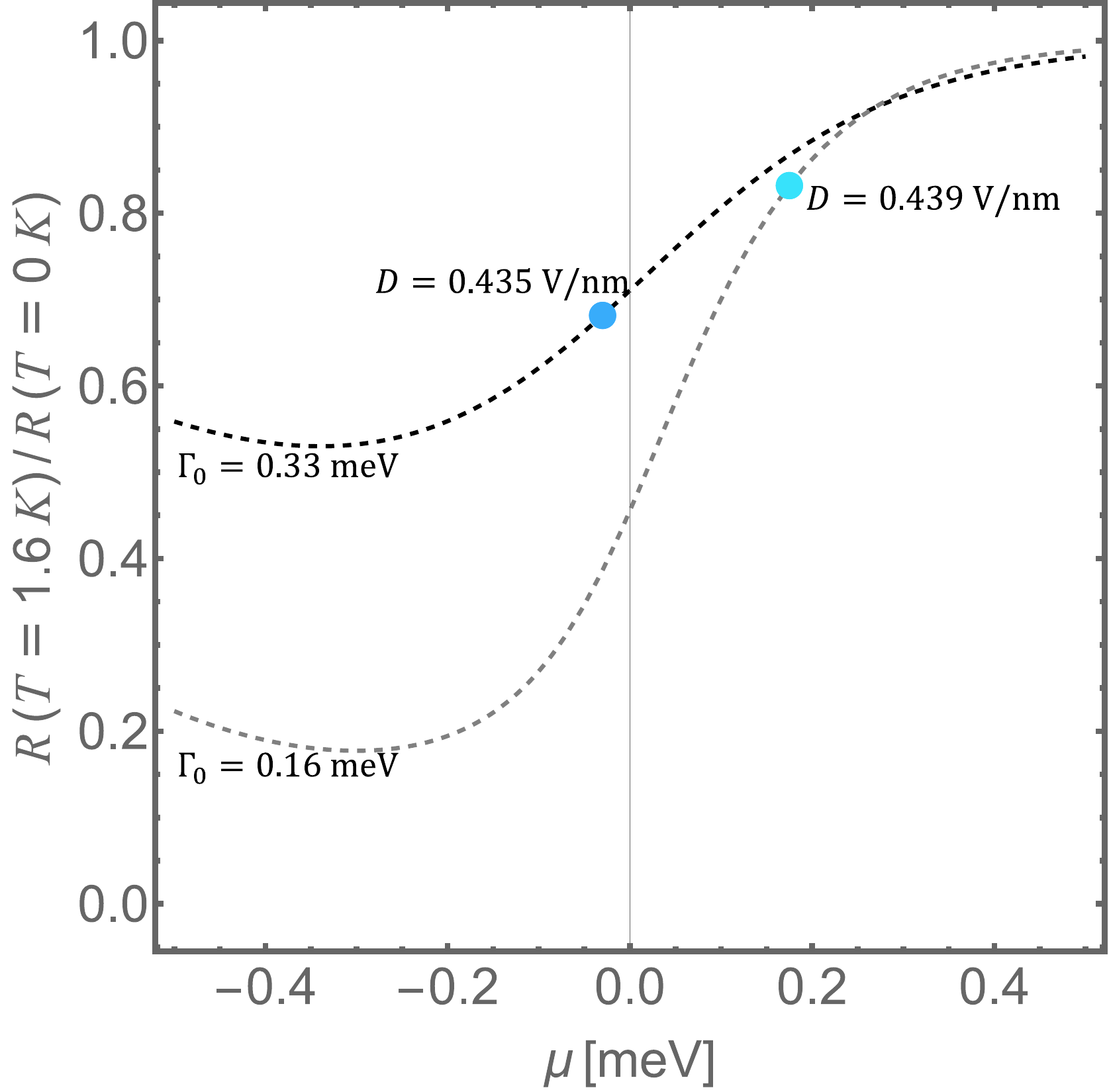}
    \caption{\textbf{Estimating $\boldsymbol{R(T=0K)}$ from $\boldsymbol{R(T=1.6K)}$.} The ratio of resistance at $T=1.6K$ and to the resistance at $T=0K$ is shown. The blue points represent the ratio for the two experimental curves (corresponding to the parameters used in our fit) which form the upper and lower bound for our estimate critical resistance range. Close to the transition point we see that $R(T=1.6K)$ is much less than $R(T=0K)$.}
    \label{extrapolating}
\end{figure}

Therefore, whilst we cannot truly verify our prediction, we can test whether our prediction is compatible with the experiment, under the assumption that the true critical curve will lie \textit{close }to the $D=0.437$ V/nm curve. Then we estimate a confidence interval for $R_c$ using our theory to extrapolate to $T=0K$. We consider the two curves which surround the $D=0.437$ V/nm curve to form the upper ($D=0.435$ V/nm) and lower ($D=0.439$ V/nm) bound for the region within which the true critical curve should lie. In order to estimate a value of $ R(T=0\text{K})$ for the experimental curves, we fitted our theory to the unscaled resistance data (shown in Fig.~\ref{direct_fit}) and by tuning the chemical potential $\mu$ and the scattering rate $\Gamma=\Gamma_0(\mu)+bT$. Note that, here, we are allowing for a chemical potential-dependent residual scattering rate to improve the precision of the fit.  Using these fits we extrapolate the value to T=0K and Fig.~\ref{extrapolating} shows how the value of  $ R(T=1.6\text{K})$ is related to the value of $  R(T=0\text{K})$. We note that, as is seen in Fig. \ref{extrapolating}, the value of $ R(T=1.6\text{K})$ is considerably less than the value of $  R(T=0\text{K})$, hence justifying why we cannot simply 'read-off' the value of $ R(T=1.6\text{K})$ to approximate the zero temperature value.

Therefore, for the MoTe$_2$/WSe$_2$ system, we get an approximate range of $ \sim11 k\Omega\leq R_c \leq \sim 47 k\Omega$. However to allow for comparison with other systems we take into account the degrees of freedom of the system, $N_d=4$ due to the 2 bands (at the transition point) and the 2 (spin-locked) degenerate valleys, by multiplying the result by 4. The critical resistance confidence range, \textit{per degree of freedom}, is therefore $ 0.3\frac{2\pi h}{e^2}\leq R_c \leq 1.2\frac{2\pi h}{e^2}$, which is the range we quote in the main text.

The critical resistance ranges for the other materials, as quoted Fig. 3 in the main text, were extracted in a very similar fashion as described for the (f=2) MoTe$_2$/WSe$_2$ system, where the two resistance curves either side of the 'critical' curve were identified and used as an upper and lower bound for $R_c$. However it should be noted that the extrapolation from finite temperature to $T=0K$ in these cases is more approximate as direct fitting to our theory was not applicable due to the non-Lifshitz transition mechanisms.

\newpage
\section{Degrees of Freedom of 2D MIT systems}

To enable a comparison between experimentally acquired values for the critical resistances in different materials,
minimal electronic structure details have to be taken into account. Namely, the number of parabolic 'bands' present. This degrees of freedom $N_d$ is decomposed into the valley degeneracy $N_v$, the band degeneracy $N_b$ and the spin degeneracy $N_s$, where $N_d=N_v\times N_b\times N_s$. The following table shows the relevant values for each system quoted in the main text.

\begin{center}
    
\begin{tabular}{|c|c|c|c|c|}
    \hline
     System&  $N_v$  & $N_b$&$N_s$&Reference\\
         \hline
 MoTe$_2$/WSe$_2$ Bilayer& 2& 2&1*&\cite{MoTe2_data} \cite{band_structure_of_MoTe2} \\ \hline 
 Monolayer WSe$_2$& 2& 1&1*&\cite{WSe2_mono_valleys}\\ \hline 

     Si& 
2  & 1&2&\cite{two_valleys_Si_MOSFET} \cite{10.1063/1.3068499}\\ \hline 
 AlAs& 1 & 1&2&\cite{alas_valleys}\cite{Al_As_data}\cite{PhysRevB.73.245214}\\ \hline 
 GaAs& 1 & 2&2&\cite{GaAs_1_valley}\\ \hline 

 SiGe& 1& 2& 2&\cite{SiGe_1}\cite{SiGe_2}\\
 \hline
 Graphene& 2& 1& 2&\cite{RevModPhys.81.109}\\ \hline
\end{tabular}
\end{center}
*Due to strong spin-orbit coupling in the valence bands of TMDs, the spin is 'locked' to the valley degree of freedom\cite{SO_1,SO_2}.


\putbib
\end{bibunit}

\end{document}